\documentclass[a4paper,fleqn,usenatbib]{mnras}

\usepackage[T1]{fontenc}
\usepackage{ae,aecompl}

\usepackage{graphicx}	
\usepackage{amsmath}	
\usepackage{amssymb}	

\newcommand{\eff}{_{\textrm{\tiny eff}}} 
\newcommand{\subsun}{_{\sun}} 

\title[The sdA problem -- III. New (pre-)ELMs from Gaia]{The sdA problem -- III. New extremely low-mass white dwarfs and their precursors from {\it Gaia} astrometry}
\author[I. Pelisoli et al.]{
	Ingrid Pelisoli$^{1,2}$\thanks{E-mail: ingrid.pelisoli@ufrgs.br},
    Keaton J. Bell$^{3,4}$,
	S. O. Kepler$^{2}$,
	D. Koester$^{5}$
	\\
	$^{1}$Institut f\"{u}r Physik und Astronomie, Universit\"{a}tsstandort Golm, Karl-Liebknecht-Str. 24/25, 14467 Potsdam, Germany\\
    $^{2}$Instituto de F\'{i}sica, Universidade Federal do Rio Grande do Sul, 91501-900, Porto-Alegre, RS, Brazil\\
    $^{3}$Max-Planck-Institut f\"{u}r Sonnensystemforschung (MPS), Justus-von-Liebig-Weg 3, 37077 G\"{o}ttingen, Germany\\
    $^{4}$Department of Physics and Astronomy, Stellar Astrophysics Centre, Aarhus University, Ny Munkegade 120, 8000 Aarhus C, Denmark\\
	$^{5}$Institut f\"{u}r Theoretische Physik und Astrophysik, Universit\"{a}t Kiel, D-24098, Kiel, Germany
}

\date{Accepted XXX. Received YYY; in original form ZZZ}

\pubyear{2018}

\begin{document}
\label{firstpage}
\pagerange{\pageref{firstpage}--\pageref{lastpage}}
\maketitle

\begin{abstract}
The physical nature of the sdA stars---cool hydrogen-rich objects with spectroscopic surface gravities intermediate between main sequence and canonical mass white dwarfs---has been elusive since they were found in Sloan Digital Sky Survey Data Release 12 spectra. The population is likely dominated by metal-poor A/F stars in the halo with overestimated surface gravities, with a small contribution of extremely low-mass white dwarfs and their precursors, i.e., ELMs and pre-ELMs. In this work, we seek to identify \mbox{(pre-)ELMs} with radii smaller than is possible for main sequence stars, allowing even for very low metallicity. We analyse 3\,891 sdAs previously identified in the Sloan Digital Sky Survey using {\it Gaia} DR2 data. Our Monte Carlo analysis supports that 90 of these are inconsistent with the main sequence. 37 lie close to or within the canonical white dwarf cooling sequence, while the remaining 53 lie between the canonical white dwarfs and main sequence, which we interpret as likely \mbox{(pre-)ELMs} given their spectral class. Of these, 30 pass more conservative criteria that allow for higher systematic uncertainties on the parallax, as well as an approximate treatment of extinction. Our identifications increase the number of known \mbox{(pre-)ELMs} by up to 50 per cent, demonstrating how Gaia astrometry can reveal members of the compact \mbox{(pre-)ELM} subpopulation of the sdA spectral class.

\end{abstract}




\section{Introduction}

The spectroscopic class of subdwarf A stars (sdAs) was proposed by \citet{dr12cat} to refer to thousands of hydrogen-rich objects in the Sloan Digital Sky Survey (SDSS) whose surface gravities derived from their low resolution spectra could not be explained by single-star evolution models. They show temperatures below the zero-age horizontal branch (ZAHB), $7\,000 \lesssim T\eff < 20\,000$~K (most $\lesssim 10\,000$~K), and $\log~g$ intermediate between main sequence A stars and hydrogen-atmosphere (DA) white dwarfs resulting from single evolution, $4.5 \lesssim \log~g \lesssim 6.0$. They were unveiled among stars identified as O, B, and A type by the SDSS pipeline. Early-type main sequence stars in SDSS are more distant than $\approx 8$~kpc to be below the saturation limit ($g \approx 14.0$), or essentially in the halo, given that SDSS observes mostly outside of the disc. As the halo is over 10~Gyr old and the main sequence lifetimes of these spectral classes is shorter than 1.5~Gyr, we do not expect a scattered large population of early-type stars in SDSS.

The initial suggestion of \citet{dr12cat} was that the sdAs could be extremely low-mass white dwarfs (ELMs) missed by the selection criteria of the ELM Survey \citep{elmsurveyI, elmsurveyII, elmsurveyIII, elmsurveyIV, elmsurveyV, elmsurveyVI, elmsurveyVII}. ELMs show masses below the single-star evolution limit for white dwarfs of $M \approx 0.3~M\subsun$ ($\log~g \approx 6.0$). The Universe is not old enough for low-mass stars to have evolved off the main sequence and turned into $M \lesssim 0.3~M\subsun$ white dwarfs \citep[e.g.][]{kilic2007}. However, over 50 per cent of stars with $M \gtrsim 1~M\subsun$ are in binaries \citep{duchene2013}, and about 25 per cent are close enough to interact \citep{willems2004}, potentially leading to a common envelope phase. The friction between the pair of stars and the envelope can lead to envelope ejection, causing significant mass loss and bringing the binary stars closer together. This is believed to be the main channel of ELM formation \citep[e.g.,][]{marsh1995,nelemans2001}.

The ELM Survey selected mostly hot ($T\eff \gtrsim 10\,000$~K), $M \gtrsim 0.15~M_\odot$ ELMs. Moreover, they considered the detection of a close binary companion as a requirement for a clear ELM classification \citep{elmsurveyVII}. Whereas common-envelope evolution is thought to be the main formation channel for ELMs, alternative formation scenarios include the merger of the inner binary in a hierarchic triple system \citep{vos2018}, supernova stripping \citep{wang2009}, and mass ejection caused by a massive planet \citep{nelemans1998}. Therefore, the existence of single ELMs should not be ruled out. The ELMs can give us important clues as to the evolution of compact binary systems and of hierarchical triple systems, which are still poorly understood \citep[e.g.][]{postnov2014,toonen2016}, hence an unbiased-sample would be a valuable asset to testing evolution and population synthesis models.

The ELM explanation for the sdAs was questioned by \citet{brown2017} and \citet{hermes2017}. \citet{hermes2017} relied on the radial velocities estimated from SDSS subspectra and concluded that the vast majority of the sdAs published by \citet{dr12cat} ($>99$ per cent) were not in close binaries, and therefore that they could not be ELMs. Other formation channels such as mergers were excluded due to the low proper motion of most of the sdAs published in \citet{dr12cat}. \citet{brown2017} suggested that sdAs were predominantly metal-poor A/F main sequence stars in the halo, arguing that the number of sdAs was far too large compared to the predicted number of cool ELMs given the density of objects estimated from the known sample. They also studied six eclipsing binaries in the sample, and found their radii to be too large for compact objects. The possibility that the objects are in the pre-ELM phase, before the stars reach the cooling branch and still show extended radii \citep[e.g.,][]{maxted2011,rappaport2015}, was not seriously considered, perhaps because the lifetimes in those phases are short. \citet{brown2017} also suggested that the use of pure-hydrogen models caused the $\log~g$ values of \citet{dr12cat} to be overestimated.

In \citet{pelisoli2018}, we showed that metallicity could not be responsible for a systematic overestimate of $\log~g$. Some values of $\log~g$ could even be underestimated by the pure-hydrogen models. Moreover, we found no dependence with metallicity for the difference between our $\log~g$ values and those estimated by the SEGUE stellar parameter pipeline \citep[SSPP,][]{lee2008}. Importantly, it became clear that the sdAs were composed of overlapping stellar populations, given a bi-modal distribution in $(g-z)$. This result was corroborated by \citet{bell2018}, who found that the sdAs show varied pulsation spectra that cannot be explained with a single population. We obtained another interesting result in \citet{pelisoli2018b} by comparing the $\log~g$ obtained from SDSS spectra with values derived from ESO X-shooter high resolution spectra. The SDSS values for the four stars observed were larger by 1~dex, which could be explained by the lack of spectral coverage below $3700$~\AA{} and low resolution in the SDSS spectra, considering that the width of the lines show little dependence on $\log~g$ for temperatures in the sdA range. This could potentially explain the inconsistency raised by \citet{brown2017}.

Clearly the SDSS spectra and colours are not enough to uniquely determine the nature of the sdA stars. They could be binary byproducts, holding the potential to shed light on the evolution of multiple systems, or main sequence stars in the halo, which could illuminate the structure and formation history of the Galaxy. Resolving their nature is therefore advantageous in both aspects. With parallax measurements from data release 2 of {\it Gaia}, we can place constraints on the radii of these objects, making their nature clearer. In this work, we explore the nature of the sdAs in light of {\it Gaia} DR~2 data. In particular, we seek to identify sdAs that must be \mbox{(pre-)ELMs} based on radius constraints from {\it Gaia} astrometry.

\section{Data Analysis}

We focused our analysis on objects from ``sample A'' of \citet{pelisoli2018}, which includes all stars with SDSS pipeline classifications of O, B, and A, and spectra with signal-to-noise ratio ($S/N$) larger than 15, for which we obtained good spectroscopic fits (38\,850 objects). Our grid covers a large range of $T\eff$ and $\log~g$ spanning from canonical mass white dwarfs to the main sequence ($6000 \leq T\eff \leq 40\,000$~K and $3.5 \leq \log~g \leq 8.0$). Metallicity in this case was taken to be solar; however, estimates of $\log~g$ or $T\eff$ were not taken into account for this selection, since the $\log~g$ estimated from SDSS spectra for this spectroscopic class proved to be unreliable \citep{brown2017,pelisoli2018,pelisoli2018b}. Hence the spectroscopic fit serves essentially as a confirmation of the spectral type. Objects for which we did not obtain a good fit were mostly giant stars whose $\log~g$ and $T\eff$ lie at the border of the grid. We also highlight objects in tables 1 and 2 of the work of \citet{pelisoli2018} in particular. The former are objects whose SDSS spectra yield $\log~g > 5.5$ when fit with solar-abundance spectral models. The latter were objects found to be more likely \mbox{\mbox{(pre-)ELMs}} than main sequence stars given a probabilistic analysis that took into account the $g-z$ colours, spatial velocities ($U, V, W$), and the fitted $\log~g$ values compared to both single \citep{bertelli2008,bertelli2009} and binary evolution models \citep{istrate2016}.

We retrieved their coordinates ($\alpha, \delta$), parallaxes ($\pi$), proper motions ($\mu_{\alpha}$ and $\mu_{\delta}$), and magnitudes in the $G$, $G_{BP}$ and $G_{RP}$ passbands from the {\it Gaia} DR2 catalogue, as well as quality control parameters, using the coordinates from SDSS and a 3'' search radius. For 490 objects there were multiple identifications within this radius; we selected the {\it Gaia} source closer to the SDSS coordinates, being left with identifications for 38\,825 objects within 3''. All but two of the unmatched sources correspond to spectra taken at the position of diffraction rays of bright stars by the SDSS. One of the remaining two could be recovered with a 5'' radius, but the other returned no match even with a 15'' radius.

\subsection{Filtering}
\label{filtering}

\begin{figure}
	\includegraphics[angle=-90,width=\columnwidth]{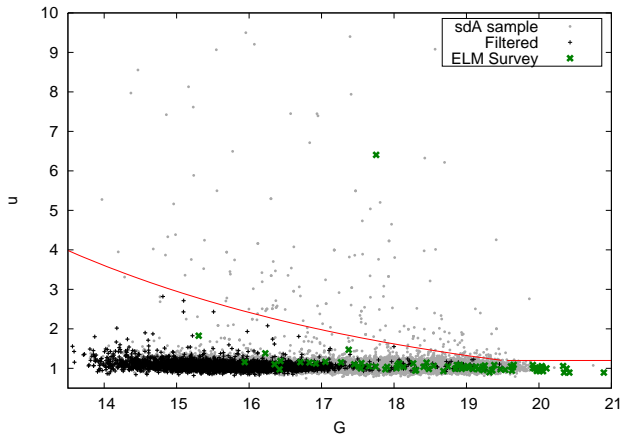}
	\caption{Accuracy of the astrometric solution unit weight $ u = \sqrt{\chi ^2 / (\nu - 5)} $ as a function of the apparent magnitude $G$ for the sdA sample (grey) and known ELMs (green crosses). The analysed sdA sample, after all filtering criteria were applied, is marked in black. The red line represents the upper limit defined by \citet{lindegren2018} for good quality data.}
	\label{chi2}
\end{figure}

Although impressive and holding potential to revolutionise astronomy, {\it Gaia} data are not perfect; thus, filtering is required to avoid contamination of spurious measurements or outliers. In order to obtain a set of filtering parameters adequate to our sample, allowing us to identify true \mbox{(pre-)ELMs} within the sdAs, we analysed the {\it Gaia} data of the 76 binary ELMs of \citet{elmsurveyVII} considering the criteria outlined in Appendix C of \citet{lindegren2018}. As can be seen in Fig.~\ref{chi2}, all the known ELMs obey the selection suggested by \citet{lindegren2018} in terms of the accuracy of the astrometric solution  unit weight error, $u = \sqrt{ \verb!astrometric_chi2_al!/(\verb!astrometric_n_good_obs_al! - 5) } $
\begin{eqnarray}
u < 1.2 \times \verb!max!(1, \exp(-0.2(\verb!phot_g_mean_mag! - 19.5)))
\end{eqnarray}
with the exception of SDSS~J093506.92+441107.0, a double WD binary with a period of 20~minutes that appears to have an M dwarf along the line of sight \citep{kilic2014} that probably affected the {\it Gaia} solution. We thus adopted the same filtering criterion as \citet{lindegren2018} on $u$.

\begin{figure}
	\includegraphics[angle=-90,width=\columnwidth]{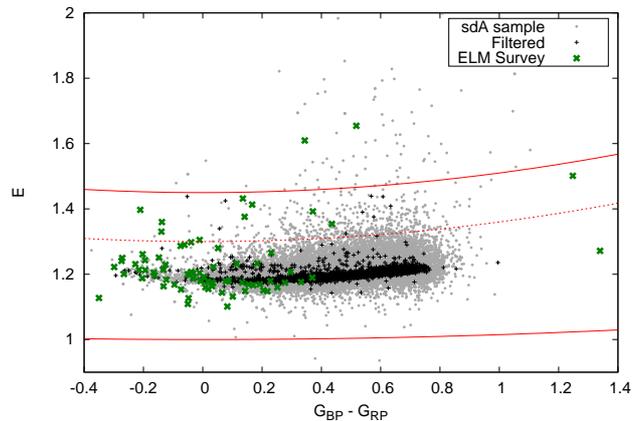}
	\caption{Flux excess factor as a function of colour, with the same colour-code as Fig.~\ref{chi2}. The solid red lines represent our adopted selection, which uses the same lower limit as \citet{lindegren2018}. The dashed line is the upper limit set by \citet{lindegren2018}, which excludes a large fraction of the known ELMs.}
	\label{excess}
\end{figure}

The next criterion suggested by \citet{lindegren2018} to obtain a clean sample relies on the flux excess factor $E = (I_{BP} + I_{RP})/I_{G}$ ($\verb!phot_bp_rp_excess_factor!$), where $I_{X}$ is the photometric flux in the band $X$ \citep{evans2018}. As shown in Fig.~\ref{excess}, the selection of \citet{lindegren2018} excludes many known ELMs (about 15~per cent). This is probably due to the fact that ELMs are binaries, so that the flux of the (cool) unseen secondary star might affect the measurement of $I_{RP}$ and cause an excess. Therefore, we relaxed the arbitrary cut of \citet{lindegren2018} to include a larger fraction of the known ELMs, adopting as a cut
\begin{eqnarray}
1.0 + 0.015(G_{BP} - G_{RP})^2 &<&\\
E &<& 1.45 + 0.06 (G_{BP} - G_{RP})^2, \nonumber
\end{eqnarray}
which includes 97 per cent of the ELMs; only SDSS~J081822.34+353618.9 and J100554.05+355014.2 are outside the selected region.

Considering the size of the sample, we further restrict our selection to include objects showing uncertainty in the parallax smaller than 25 per cent, i. e. $\verb!parallax_over_error! > 4$, excluding about 90 per cent of our initial sample \citep[sample A of ][]{pelisoli2018}. Assuming Gaussian uncertainties, this 4-$\sigma$ cut would suggest that fewer than three stars in our analysed sample have spurious parallax detections. Although it is likely that the {\it Gaia} distribution of parallax outliers is non-Gaussian, outliers are concentrated in dense regions, such as the Galactic bulge and Galactic plane \citep{gaia2018}. As our sources come from SDSS, which observes outside of these regions and in the Northern hemisphere, avoiding also contamination by the Magellanic clouds, there is no reason to believe a significant number of our analysed sources could be outliers, especially given that the selection criteria of \citet{lindegren2018} were taken into account, as outlined above. Moreover, parallax outliers likely show also erroneous proper motions. We have cross-checked the {\it Gaia} proper motion of our sources, especially those found to be below the main sequence, with previous determinations and found no discrepancy (see Appendix~\ref{extra}). We caution that a similar 4-$\sigma$ cut in the sample of known ELMs would exclude more than 60 per cent of the objects.

\begin{figure}
	\includegraphics[angle=-90,width=\columnwidth]{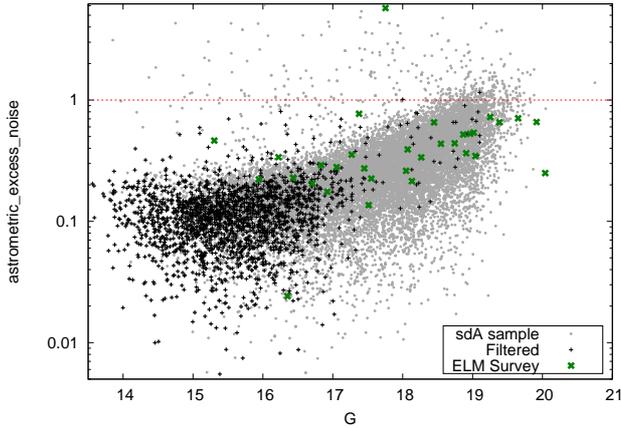}
	\caption{Astrometric excess noise as a function of magnitude. Colours for the symbols are the same as in Fig.~\ref{chi2}. The dashed line marks the position of astrometric excess noise equal to 1~mas.}
	\label{noise}
\end{figure}

The outlined selection criteria have also excluded from our sample objects with high $\verb!astrometric_excess_noise!$, a parameter that quantifies how much the solution deviates from the single-source solution adopted by {\it Gaia}. The binarity of the ELMs is of course expected to cause deviations; however, as the binaries are too close to be resolved by {\it Gaia}, they should still be relatively well modelled by a single-star solution, as can be noted from Fig.~\ref{noise}. The only known ELM that shows high noise is SDSS~J093506.92+441107.0, which was also an outlier in the $u$ selection.

\begin{figure}
	\includegraphics[width=\columnwidth]{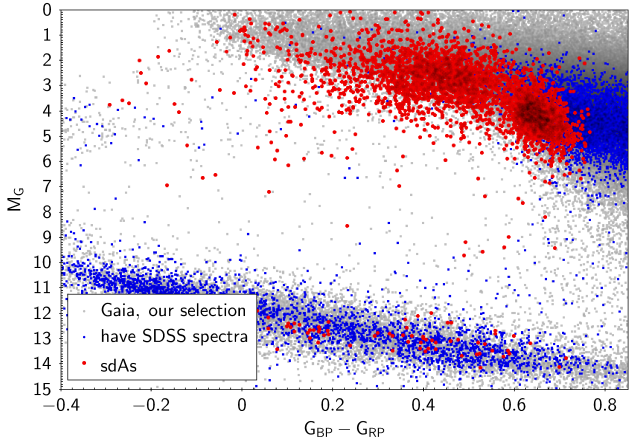}
	\caption{{\it Gaia} HR diagram of objects in SDSS~DR12 (grey), with those for which spectra were obtained shown in blue. The sample analysed here is shown in red, and corresponds to the region where spectra are dominated by hydrogen lines.}
	\label{sdss_hr}
\end{figure}

Removing double identifications (objects with more than one observation in the SDSS, with slightly different coordinates), we are left with 3\,891 unique identifications obeying our selection, which we will refer to as the ``reliable $\pi$'' sample throughout the text. In Fig.~\ref{sdss_hr}, we show the position in the HR diagram of these objects compared to other SDSS targets that follow the selection criteria outlined above. This illustrates that our selection corresponds to objects in the region where the spectra show only H lines. The few objects with SDSS spectra that are not in our sample have $S/N$ lower than 15, sometimes leading to an erroneous classification by the SDSS pipeline. It can also be seen that the SDSS sample of objects with observed spectra (in blue) shows less scatter than the full sample (in grey), with the same selection parameters, i.e. focusing on objects with spectra reduces the chances of outliers. We note that this subsample is biased to include smaller stars at a given magnitude, as we expect nearer sources to have higher signal-to-noise parallax measurements.

\subsection{Radii estimates}
\label{sec:radiiest}

We estimate the radii of the sdAs that are required to bring the observed magnitudes, effective temperatures, and parallaxes into agreement, largely following the approach of \citet{andrae2018}, with a few improvements. First, we have obtained effective temperatures by fitting SDSS spectra and selecting the solution that best agrees with SDSS and GALEX photometry, allowing for a more accurate determination than that solely based on {\it Gaia} photometry as done by \citet{andrae2018}. Moreover, we propagate uncertainties in all variables via Monte Carlo, rather than the simple first-order approximation adopted by \citet{andrae2018}, which also neglects the effect of the luminosity uncertainty in the radius. The steps of our estimates are described in more detail below.

We subtracted the zero-point of $-29~\mu$as from the parallaxes \citep{lindegren2018} of all matched objects. This zero-point was determined from quasars, which have similar colours to white dwarfs and A-type stars, hence it should be adequate for our sample. There is, however, a dependence of the magnitude of this zero-point with colour, and it can be larger for cooler objects such as the sdAs \citep[see e.g.][]{zinn2018}. A greater negative zero-point would imply larger parallax, which in turn implies smaller distance and therefore smaller radius. Thus our assumption of a $-29~\mu$as zero-point could cause our radii to be slightly overestimated. As we are searching the objects with radii smaller than main sequence, this can be considered a conservative zero-point.

We estimated the distance $d$, absolute magnitude $M_G$, bolometric magnitude $M_{\textrm{bol}}$, luminosity $\log (L/L\subsun)$, and radius $R/R\subsun$ for each object with a Monte Carlo simulation, taking into account the uncertainties in the value of the parallax $\pi$. The uncertainties in the $G$ magnitude, and in $T\eff$ \citep[initially assumed to be equal to the external uncertainty of 5 per cent;][]{pelisoli2018,pelisoli2018b} are also propagated. Expected values were taken to be the medians of each distribution, whereas upper and lower uncertainties were taken to be the $16th$ and $84th$ percentiles, yielding an interval of confidence of 68 per cent.

The distance $d$ for each individual source was estimated from the distribution of $(\pi')^{-1}$, where $\pi'$ is the parallax corrected for the zero point---an improvement compared to the simple inversion of parallax employed by \citet{andrae2018}. We opted not to use any priors due to the fact that, as shown by \citet{pelisoli2018} and \citet{bell2018}, the sdAs are composed by multiple populations, hence no single prior would be adequate to the whole sample. This choice can somewhat bias the distance estimate, especially for distant stars ($d \gtrsim 2$~kpc), whose distribution might show an extended tail to unphysically large distances \citep{bailerjones2015,luri2018}. However, the impact of this is mitigated by the fact that we only analyse objects with small fractional parallax uncertainty ($\sigma_{\pi}/\pi < 0.25$). In fact, our distance estimates show very good agreement up to $\approx 1.5$~kpc with the estimates of \citet{bailerjones2018}, who assumed an exponentially decreasing space density prior in distance. Above 1.5~kpc, our distances are somewhat larger, which will cause larger radii estimates compared to what we would obtain by using a single prior. This choice again errs on the side of caution in our effort to identify \mbox{(pre-)ELMs} among the sdA stars.

\begin{figure}
	\includegraphics[width=\columnwidth]{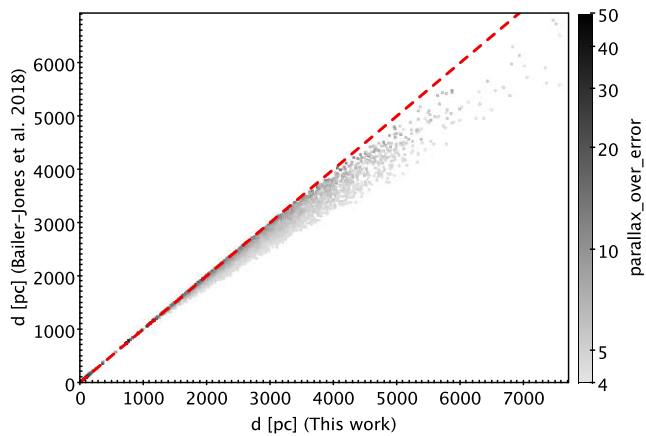}
	\caption{Comparison between our estimated distances and those of \citet{bailerjones2018}, colour-coded by the parallax over error parameter. The dashed red line represents equality. For nearby objects ($ d \lesssim 1.5$ ~kpc), there is remarkable agreement. Above this, our distances are somewhat larger than those of \citet{bailerjones2018}.}
	\label{dbailerjones}
\end{figure}

\begin{figure}
	\includegraphics[angle=-90,width=\columnwidth]{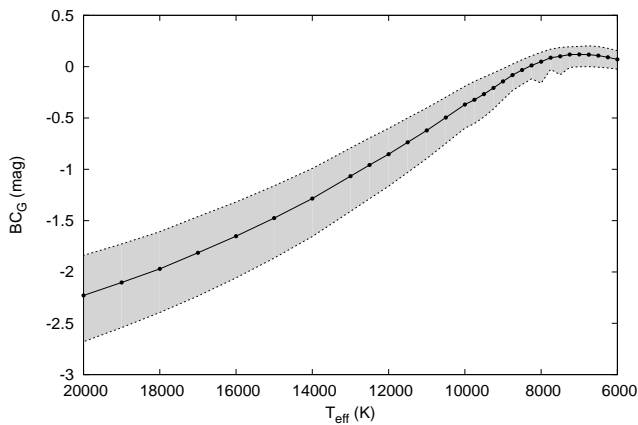}
	\caption{The black points show the average values of the bolometric correction to $M_G$ for each $T\eff$ \citep{jordi2010}. They give a smooth function represented by the solid line, hence the bolometric correction can be estimated from interpolation given the $T\eff$. The dashed lines represent the minimum and maximum value for each $T\eff$, given the variations with $\log~g$ and metallicity.}
	\label{BC}
\end{figure}

The absolute magnitude for each source was calculated as
\begin{eqnarray}
	M_G &=& G + 5\log\pi' + 5.
\end{eqnarray}
We do not account for extinction at this moment, given that the extinction in the {\it Gaia} archive is not recommended for the analysis of stars individually \citep{andrae2018}. The effect of neglecting extinction is discussed below. We applied the bolometric corrections given in table 12 of \citet{jordi2010} to obtain $M_{\textrm{bol}}$. As the external uncertainty in the $\log~g$ that we derived from SDSS spectra is quite high \citep[about 0.5~dex, see discussion in][]{pelisoli2018b} and we do not have high resolution spectra to accurately determine the metallicity, we assume the average bolometric correction given the effective temperature \citep[which we estimated to have an uncertainty smaller than 5 per cent when derived from the SDSS spectra, see][]{pelisoli2018,pelisoli2018b}, with the uncertainty taken to be the difference between either the minimum or the maximum value and the average --- whichever was larger. As can be seen in Fig.~\ref{BC}, the bolometric correction does not have a large dependence on $\log~g$ and metallicity in the temperature range of the sdAs, so the uncertainties from this method are only on the order of 0.5~mag, still they were also propagated in our simulations. Using the bolometric magnitude, we calculated $\log (L/L\subsun)$ from
\begin{eqnarray}
\log\left( \frac{L}{L\subsun} \right) &=& \frac{4.74 - M_{\textrm{bol}}}{2.5}.
\end{eqnarray}

The radius of each object was finally obtained by using the Stefan-Boltzmann law relating $L$, $T\eff$ and $R$. The main sources of uncertainty in this method are the bolometric correction and the extinction. 

To estimate the effect of the bolometric correction, we calculated a second set of radii estimates independent of the bolometric correction, by using the solid angle obtained as part of our photometric fit to each object. As the spectroscopic fit can show degenerate solutions with similar $\chi^2$, we first fit the photometry with stellar models that assumes a fixed $\log~g = 4.5$ so that we can select the spectroscopic solution whose temperature is consistent with the photometry. The ratio between the observed and modelled photometric brightness is related through the solid angle, which is proportional to $(R/d)^2$. Hence, the radius can be estimated given the distance. In this case the main uncertainty is only the extinction correction. We applied the full-correction from the maps of \citet{schlegel1998}, which have a resolution of only 4', to the $ugriz$ magnitudes previous to our photometric fit. The radius estimates from the two methods were not consistent within $3\sigma$ for only 53 objects (1.4 per cent), suggesting that the bolometric correction has no large effect. Henceforth we use the estimate from the luminosity, which is independent from our fit to the SDSS photometry.

Neglecting the extinction correction can cause the distances and radii to be underestimated, a major concern given that we are interested in finding the objects with radii smaller than main sequence. To estimate the effect of a lack of extinction correction, we made a third calculation of radii obtaining the apparent Gaia magnitude $G$ from the SDSS filters $g$ and $i$ using the transformations of \citet{evans2018}, and correcting extinction in both SDSS filters following \citet{schlegel1998}. We consider how this correction affects our \mbox{(pre-)ELM} identifications in Section~\ref{sec:results}.

Another concern is the systematic uncertainty in the {\it Gaia} parallax, which is about 0.1~mas \citep{luri2018}. To verify the effect of this on our radii estimates, we validate those identifications of \mbox{(pre-)ELMs} that remain inconsistent with the main sequence after quadratically adding this to the uncertainties reported in the catalogue.

Given the {\it Gaia} five-parameter astrometric solutions, we have also computed the galactic coordinates $X, Y, Z$ and spatial velocities $U, V, W$ for each of our targets following \citet{johnson1987}. Given that they are too faint to have radial velocity estimates from {\it Gaia}, we have relied on the radial velocities that we estimated from their SDSS spectra. Our estimates derived from the hydrogen lines show very good agreement with the estimates from the SDSS pipeline, within an average uncertainty of 20 km/s.

The radial velocity variations between SDSS subspectra of the objects in this sample were also analysed. These amplitudes were calculated with the approach described in \citet{pelisoli2017}, which was based on the work of \citet{badenes2012}. It consists of fitting a Gaussian to each of the Balmer lines in the normalized spectra to determine the line centre, then using this value to compute a radial velocity for each line. The radial velocity of the object at the epoch of each spectrum is calculated to be the average velocity over all the lines. We then calculate the difference between minimal and maximal velocities for each object, which is a proxy for the amplitude of radial velocity variation. We obtained good fits to the subspectra of 3\,664 objects in the sample.

\section{Results}
\label{sec:results}

\begin{figure}
	\includegraphics[angle=-90,width=\columnwidth]{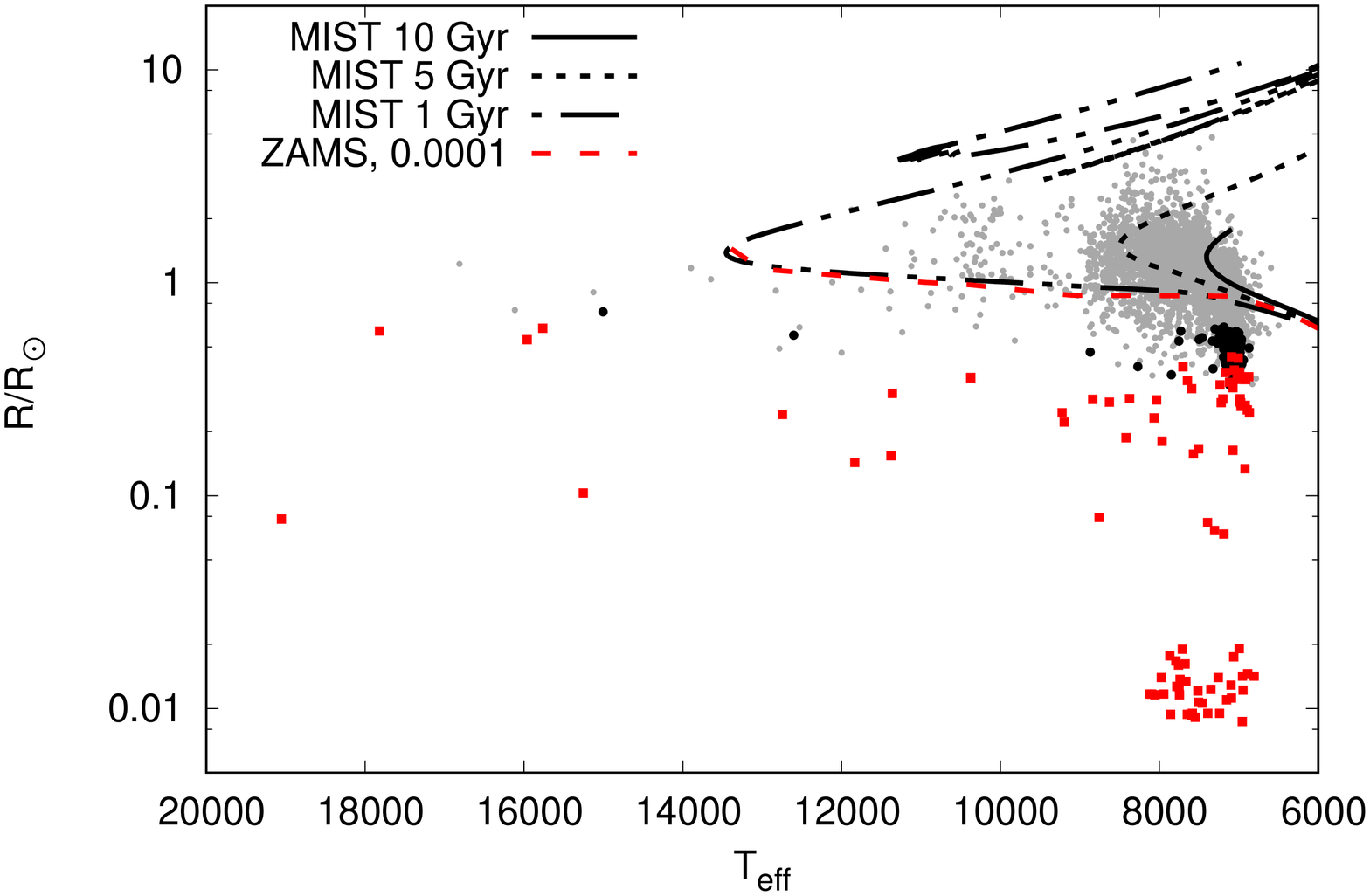}
	\caption{Radius as a function of $T\eff$ for all objects with reliable parallax. The red long-dashed line shows the zero-age main sequence (ZAMS) for the indicated metallicity \citep{romero2015}. The black lines show the radii from MIST isochrones \citep{mist1,mist2} for [Fe/H] = $-4$. The objects showing radii smaller than the minimum main sequence radius, taking into account a 99 per cent confidence interval and assuming a 5 per cent uncertainty in $T\eff$, are shown as black points. Red squares are the objects which are still below the main sequence when 10 per cent in $T\eff$ uncertainty is assumed.}
	\label{radii}
\end{figure}

The main candidate explanations put forward for the sdAs are A/F main sequence stars in the halo \citep{brown2017}, and \mbox{(pre-)ELMs} \citep{pelisoli2018}. The main difference between these two evolutionary classes is their radii. A/F main sequence stars show $R \gtrsim 0.6~R\subsun$, depending on $T\eff$, even for very low metallicity (e.g. $z \approx 10^{-4}$, see Fig.~\ref{radii}). With larger metallicity, the radius is larger because of the increase in opacity. Evolution off the main sequence also causes the radius to increase, due to thermal energy being released by the contracting nucleus. Hence, the zero age main sequence (ZAMS) radius for low metallicity is the conservative minimal radius for a main sequence star at a given $T\eff$. \mbox{(Pre-)ELMs}, on the other hand, have $R \lesssim 0.1~R\subsun$ during most of their evolution. The radius of a pre-ELM can nonetheless be larger because of residual burning \citep[e.g., the RR~Lyra found by][]{pietrzynski2012}. Although the time spent in these burning stages is short, the objects are much brighter, hence they have considerable chances of being detected \citep{pelisoli2018}. In short, \mbox{(pre-)ELMs} can show radii in an extensive range, but main sequence stars have a minimal radius. 

We have used this minimal radius to investigate which objects in our sample could not be explained as main sequence stars. Using the radii estimated from the parallaxes, we verified whether each star was smaller than the ZAMS at the estimated $T\eff$ considering the 99 per cent confidence interval of our Monte Carlo simulation---i.e. 99 per cent of the simulated radii had to be below the minimum radius at the best-fit $T\eff$ for it to be considered smaller than the ZAMS. From the sample of 3\,891 objects, we found 234 showing $R_{99\%} < R_{\textrm{ZAMS}}$, highlighted in Fig.~\ref{radii}. Objects with radii below the ZAMS cannot be main sequence stars. Removing canonical mass white dwarfs with spectral types other than pure A (e.g. with helium contamination or magnetic fields, which affect the $\log~g$ estimate), we are left with 187 objects whose radii estimate places them below the main sequence (see Fig.~\ref{radii}).

However, with our assumption of 5 per cent uncertainty in $T\eff$, a fraction of these 187 objects are clustered at $T\eff \approx 7\,000$~K (see Fig.~\ref{radii}). The binary evolution models indicate no physical reason for this clustering. Although comparison with other temperature estimates, e.g. from \textit{Gaia} \citep{andrae2018}, suggest uncertainties of, at most, 6.5 per cent in our $T\eff$, at lower temperatures the external uncertainty of the models does become larger, due to increased uncertainty in the opacities. We hence decided to adopt a more conservative uncertainty of 10 per cent. The slope of minimal radius along the main sequence steepens for $T\eff < 7000$~K, so that this increase in the uncertainty allows most of the objects in the clustered region to show radii consistent with the cooler main sequence considering the 99 per cent confidence level. 90 objects remain below the main sequence, two of which are known ELMs (SDSS~J123800.09+194631.4 and SDSS~J155502.00+244422.0).

\begin{figure}
	\includegraphics[width=\columnwidth]{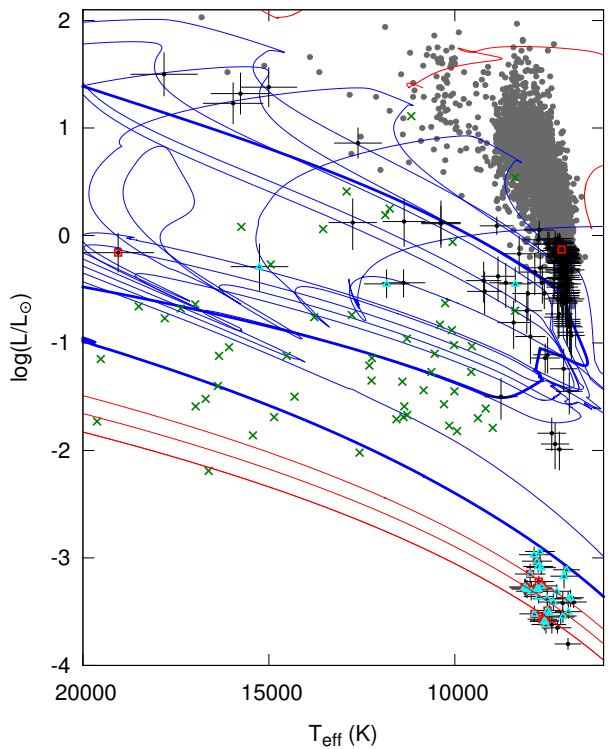}
	\caption{H-R diagram showing the whole reliable parallax $\pi$ sample as grey dots. The objects with radii smaller than main sequence assuming a 99 per cent confidence interval and 10 per cent uncertainty in $T\eff$ are shown in black, with error bars. Objects found to be most likely \mbox{(pre-)ELMs} from \citet{pelisoli2018} are marked with cyan triangles, while those with $\log~g > 5.5$ are marked with red squares. Blue lines show the binary evolution models of \citet{istrate2016} for $z=0.01$ that take rotation into account, ranging from $0.182$ (thin) to $0.324~M\subsun$ (thick). The location of the \mbox{(pre-)ELMs} agrees well with the region spanned by these models, with the exception of a few canonical-mass white dwarfs. Known ELMs in binaries from table 5 of \citet{elmsurveyVII} are shown as green crosses for comparison. The red lines are single-evolution models computed with the {\sc LPcode} \citep{althaus2003} for low-metallicity ($z=0.004$) stars with initial masses of 1.0, 2.0, and 3.0~$M\subsun$ (from top to bottom).}
	\label{HR}
\end{figure}

In Fig.~\ref{HR}, we show the H-R diagram for our reliable parallax $\pi$ sample. Single and binary evolution models are overplotted. It is evident that $\approx 40$ per cent of the objects with radii below the main sequence (37 stars) are consistent with the single-evolution models of canonical mass white dwarfs. These objects are all included in the {\it Gaia} white dwarf catalogue of \citet{nicola2018}. Most were previously published in SDSS catalogues \citep{dr7cat,dr10cat,dr12cat}, except for two (SDSS~J031637.81-003310.9 and SDSS~J214412.22+092630.0). Nine have estimated masses below $0.45~M\subsun$, being classified as low-mass white dwarfs ($0.3 < M < 0.45~M\subsun$), which are also likely found in binaries like the ELMs, but show a much smaller binarity rate (30 per cent), suggesting single-evolution channels are also required \citep{jbrown2011}. The remaining have masses estimated to be canonical ($M > 0.45~M\subsun$), with the narrow hydrogen lines being caused likely by low $T\eff$. We note that some of the known ELMs in tight binaries also lie in the region of the single-evolution models and still show low estimated masses (e.g. J233821.504$-$205222.78, with a mass of $0.258~M\subsun$, and J115138.378+585853.20, with a mass of $0.186~M\subsun$).

\begin{figure}
	\includegraphics[width=\columnwidth]{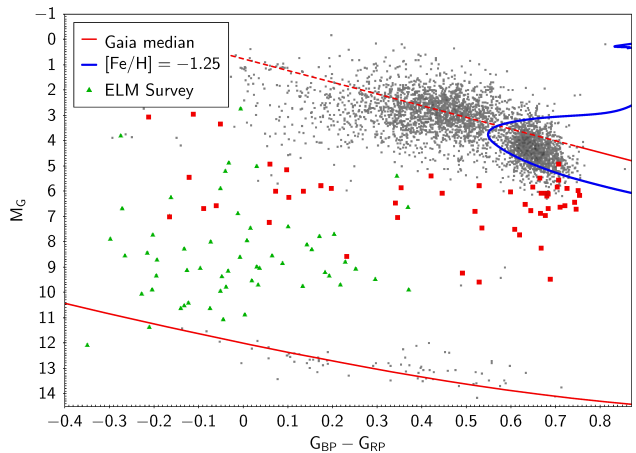}
	\caption{Absolute magnitude $M_G$ and colour $G_{BP}-G_{RP}$ for the reliable $\pi$ sample as grey dots. The objects with radii smaller than main sequence assuming a 99 per cent confidence interval and 10 per cent uncertainty in $T\eff$ are marked with red squares. The red lines are the fiducial lines from fig.~8 of \citet{babusiaux2018}, which show the position of the main sequence (top) and white dwarfs (bottom). The dashed portion was extrapolated. The blue line is the MIST isochrone for 10~Gyr and [Fe/H] = $-1.25$.}
	\label{MgBp}
\end{figure}

There remain thus 53 objects whose radii is smaller than the minimal radius for main sequence stars, and yet are inconsistent with white dwarfs resulting from single-evolution models. Considering their spectral class, these are high-probability \mbox{(pre-)ELMs}. These 53 targets are listed in Table~\ref{preELMs}, except for three previously known ELMs published by \citet{elmsurveyVII} (SDSS~J123800.09+194631.4 and SDSS~J155502.00+244422.0) or by \citet{pelisoli2018b} (SDSS~J142421.30-021425.4). Their position in Fig.~\ref{HR} agrees quite well with the binary evolution models of \citet{istrate2016}, providing further evidence that they are, in fact, ELMs and their precursors. These objects also lie below the main sequence in the observational H-R diagram (colour-magnitude diagram), shown in Fig.~\ref{MgBp}.

Only one out of the 53 \mbox{(pre-)ELMs} shows inconclusive radii in our simulations that approximate extinction based on the SDSS magnitudes, SDSS~J062219.10+001003.7. On the other hand 21 objects ($\approx 40$~per cent) have inconclusive radius when the systematic parallax uncertainty is added in quadrature to the catalogue uncertainties (see Section~\ref{sec:radiiest}); they are flagged in Table~\ref{preELMs}.

\begin{table*}
	\centering
	\caption{The 50 new high probability \mbox{(pre-)ELMs} identified in the sdA sample. Uncertainties in $T\eff$ and $\log~g$ are formal fitting errors obtained from the SDSS spectral fit. The radii were calculated from the luminosity derived from {\it Gaia} data and our spectroscopic $T\eff$. The flag $F_{\textrm{sys}}$ is 0 when the object has inconclusive radius when the systematic uncertainty in parallax is added in quadrature to the catalogue uncertainties.}
	\label{preELMs}
	\begin{tabular}{ccccccccccc}
		\hline
		\hline
		SDSS~J & $T\eff$ (K) & $\log g$ & $d$ (pc) & $\sigma_{d\textrm{\tiny upper}}$ & $\sigma_{d\textrm{\tiny lower}}$ & $R (R\subsun)$ & $\sigma_{R\textrm{\tiny upper}}$ & $\sigma_{R\textrm{\tiny lower}}$ & $F_{\textrm{sys}}$ & SDSS P-M-F \\
		\hline
002543.80-104706.8 &  6919( 35) &  3.87( 0.18) & 1529 & 447 & 286 & 0.2640 & 0.0855 & 0.0559 & 0 & 1912-53293-0097 \\ 
012000.11+003251.4 &  7019( 14) &  4.21( 0.08) & 1393 & 267 & 193 & 0.3494 & 0.0820 & 0.0601 & 1 & 1080-52614-0507 \\ 
020148.52+055355.0 &  8375(  9) &  4.89( 0.04) & 1488 & 305 & 219 & 0.2860 & 0.0697 & 0.0507 & 1 & 3116-54792-0217 \\ 
062219.10+001003.7 & 10374( 97) &  4.42( 0.04) & 2234 & 624 & 388 & 0.3560 & 0.1163 & 0.0770 & 0 & 1259-52931-0083 \\ 
073553.29+441314.9 &  8035( 16) &  5.06( 0.07) & 2007 & 618 & 382 & 0.2781 & 0.0970 & 0.0619 & 0 & 3225-54853-0334 \\ 
075246.16+175342.6 &  7504( 24) &  4.53( 0.10) &  980 & 197 & 141 & 0.1648 & 0.0412 & 0.0303 & 1 & 1921-53317-0444 \\ 
080652.41+455326.4 &  7238( 12) &  4.31( 0.04) & 1645 & 338 & 239 & 0.3265 & 0.0796 & 0.0584 & 0 & 3686-55268-0234 \\ 
085332.85+385050.9 &  7393( 30) &  4.38( 0.19) &  568 & 102 &  75 & 0.0741 & 0.0170 & 0.0128 & 1 & 1198-52669-0227 \\ 
085429.85+031447.3 &  7189( 22) &  4.02( 0.11) &  679 & 161 & 108 & 0.0660 & 0.0180 & 0.0127 & 1 & 2913-54526-0486 \\ 
085745.55+203850.4 &  8759( 42) &  4.29( 0.11) &  834 & 225 & 142 & 0.0778 & 0.0235 & 0.0159 & 1 & 2282-53683-0134 \\ 
091753.19+523004.8 &  6987( 15) &  4.19( 0.08) & 1365 & 166 & 131 & 0.3698 & 0.0639 & 0.0535 & 1 & 0553-51999-0614 \\ 
094400.28+620224.2 &  6874( 23) &  3.88( 0.07) & 1723 & 343 & 252 & 0.3601 & 0.0851 & 0.0646 & 0 & 2383-53800-0172 \\ 
095813.66+314106.9 &  7304( 26) &  4.49( 0.10) &  731 & 221 & 136 & 0.0676 & 0.0230 & 0.0142 & 1 & 6468-56311-0476 \\ 
105101.41-003523.9 &  7117( 16) &  4.00( 0.09) & 1663 & 416 & 282 & 0.3370 & 0.0985 & 0.0656 & 0 & 2389-54213-0107 \\ 
111129.23+203559.3 &  7093( 11) &  4.08( 0.06) & 1429 & 167 & 133 & 0.4466 & 0.0775 & 0.0620 & 0 & 2492-54178-0208 \\ 
112157.13+605210.4 & 11836( 31) &  5.29( 0.01) &  751 &  29 &  28 & 0.1427 & 0.0279 & 0.0233 & 1 & 3328-54964-0149 \\ 
112914.16+471501.7 & 11379( 49) &  5.01( 0.02) &  814 &  47 &  43 & 0.1552 & 0.0297 & 0.0246 & 1 & 3329-54970-0270 \\ 
113321.07+451747.6 &  6983( 17) &  4.25( 0.09) & 1624 & 463 & 299 & 0.2799 & 0.0912 & 0.0582 & 0 & 3215-54861-0501 \\ 
122021.90+002249.3 &  7223( 16) &  4.33( 0.07) & 1313 & 331 & 223 & 0.2687 & 0.0774 & 0.0516 & 1 & 2558-54140-0428 \\ 
124033.34+231633.3 &  6922( 31) &  4.05( 0.17) & 1016 & 271 & 179 & 0.1327 & 0.0397 & 0.0275 & 1 & 3374-54948-0066 \\ 
124357.97+385646.1 &  8630( 21) &  4.31( 0.07) & 2051 & 591 & 370 & 0.2729 & 0.0888 & 0.0566 & 0 & 2000-53495-0584 \\ 
124451.71-015109.0 &  7165( 18) &  4.13( 0.09) & 1585 & 307 & 222 & 0.3773 & 0.0887 & 0.0658 & 0 & 2897-54585-0536 \\ 
124656.44+304238.0 &  6989( 19) &  4.22( 0.10) & 1498 & 308 & 215 & 0.2738 & 0.0681 & 0.0481 & 1 & 2457-54180-0537 \\ 
125328.45+042044.0 & 12745( 88) &  3.54( 0.03) & 1779 & 478 & 319 & 0.2377 & 0.0825 & 0.0576 & 1 & 0847-52426-0021 \\ 
125859.32+252731.5 &  7057( 14) &  4.25( 0.05) & 1371 & 191 & 149 & 0.3883 & 0.0725 & 0.0589 & 1 & 2662-54505-0324 \\ 
130953.20+383816.0 &  6916( 28) &  4.01( 0.15) & 1933 & 461 & 315 & 0.3492 & 0.0938 & 0.0672 & 0 & 2900-54569-0204 \\ 
131533.95+252211.7 &  7703( 13) &  4.64( 0.07) & 2013 & 379 & 263 & 0.4032 & 0.0919 & 0.0684 & 0 & 3303-54950-0389 \\ 
132446.98+062937.0 &  6974( 18) &  4.13( 0.10) &  921 & 100 &  82 & 0.2617 & 0.0426 & 0.0358 & 1 & 1799-53556-0169 \\ 
132654.55+122912.5 &  8066( 23) &  4.29( 0.10) & 1610 & 413 & 262 & 0.2300 & 0.0678 & 0.0469 & 1 & 1698-53146-0581 \\ 
132713.01+382514.0 &  7967( 18) &  4.58( 0.10) & 1407 & 434 & 272 & 0.1790 & 0.0626 & 0.0398 & 1 & 3240-54883-0068 \\ 
134326.35+205914.3 &  6993( 18) &  4.18( 0.10) & 1377 & 167 & 133 & 0.3801 & 0.0659 & 0.0542 & 1 & 2654-54231-0079 \\ 
140158.86+181427.1 & 11362( 46) &  4.87( 0.02) & 2311 & 602 & 400 & 0.3001 & 0.0999 & 0.0671 & 0 & 3310-54919-0550 \\ 
141137.24+585928.2 &  8839( 23) &  4.46( 0.07) & 2315 & 681 & 428 & 0.2779 & 0.0914 & 0.0594 & 0 & 0788-52338-0180 \\ 
141730.36+002601.2 &  7074( 17) &  4.24( 0.09) & 1440 & 218 & 171 & 0.3234 & 0.0635 & 0.0502 & 1 & 0304-51957-0386 \\ 
142311.27+573103.4 &  6865( 34) &  4.06( 0.13) & 1587 & 434 & 286 & 0.2432 & 0.0751 & 0.0498 & 1 & 2547-53917-0282 \\ 
150422.91+002731.7 &  6893( 39) &  3.96( 0.21) & 1467 & 392 & 260 & 0.2504 & 0.0755 & 0.0511 & 1 & 0539-52017-0086 \\ 
151500.55+072548.5 &  6976( 32) &  4.00( 0.17) & 1569 & 396 & 267 & 0.2677 & 0.0767 & 0.0530 & 1 & 2724-54616-0426 \\ 
151552.17+065452.1 &  7073( 23) &  4.20( 0.12) & 1197 & 321 & 214 & 0.1609 & 0.0492 & 0.0328 & 1 & 2739-54618-0189 \\ 
152645.40+515002.5 & 10373( 74) &  3.88( 0.03) & 1845 & 369 & 256 & 0.3559 & 0.0923 & 0.0688 & 0 & 0795-52378-0255 \\ 
160502.57+453831.1 &  7010( 16) &  3.98( 0.09) & 2050 & 431 & 293 & 0.3597 & 0.0871 & 0.0638 & 0 & 3428-54979-0388 \\ 
165109.68+390831.2 &  7008( 14) &  4.46( 0.05) & 1664 & 173 & 141 & 0.4455 & 0.0714 & 0.0608 & 0 & 0630-52050-0048 \\ 
170716.53+275410.3 &  7594( 11) &  4.70( 0.05) & 1379 & 222 & 166 & 0.3131 & 0.0655 & 0.0513 & 1 & 2808-54524-0069 \\ 
172720.78+324908.4 & 15763( 90) &  4.02( 0.02) & 2604 & 306 & 243 & 0.6139 & 0.1619 & 0.1297 & 0 & 2253-54551-0307 \\ 
192242.87+634618.4 & 17819(149) &  4.16( 0.03) & 3058 & 447 & 352 & 0.5889 & 0.1732 & 0.1299 & 0 & 2553-54631-0403 \\ 
200400.51-102112.6 & 15961(106) &  4.07( 0.02) & 2340 & 322 & 246 & 0.5388 & 0.1486 & 0.1146 & 0 & 2303-54629-0366 \\ 
220919.19-005734.0 &  7647( 18) &  4.62( 0.11) & 1801 & 420 & 290 & 0.3432 & 0.0929 & 0.0659 & 0 & 3146-54773-0256 \\ 
221901.67-082259.1 &  7570( 16) &  4.51( 0.08) &  834 & 125 &  95 & 0.1563 & 0.0327 & 0.0251 & 1 & 0719-52203-0595 \\ 
223532.78+005955.6 &  7199( 33) &  4.25( 0.17) & 1358 & 386 & 249 & 0.2814 & 0.0883 & 0.0595 & 1 & 1101-52621-0333 \\ 
230911.41-000433.9 &  9224( 16) &  4.52( 0.03) & 1487 & 342 & 234 & 0.2427 & 0.0676 & 0.0485 & 1 & 0381-51811-0200 \\ 
233255.72+520431.3 &  8419( 42) &  3.54( 0.10) & 1786 & 551 & 343 & 0.1864 & 0.0609 & 0.0414 & 1 & 1662-52970-0009 \\ 
\hline
	\end{tabular}
\end{table*}

From the 1150 objects identified as most likely \mbox{(pre-)ELMs} by \citet{pelisoli2018}, 150 were in our reliable $\pi$ sample, and 54 were found to indeed show radii smaller than the main sequence; most are, however, consistent with single evolution models. As to the 411 objects with $\log~g > 5.5$ from fits to their SDSS spectra, 94 were analysed and we established that 20 must be below the main sequence. Many are possibly canonical mass white dwarfs, considering their position in the H-R diagram of Fig.~\ref{HR}. Given our conservative criteria, we cannot rule out that individual remaining objects are \mbox{(pre-)ELMs}, and the identified \mbox{(pre-)ELMs} represent an unknown fraction of this subpopulation of sdA stars. In Appendix~\ref{extra}, we show a comparison between the distances and proper motions quoted in the work of \citet{pelisoli2018} compared with the updated values from {\it Gaia}.

\section{Discussion}

\subsection{A halo population of cool subdwarfs}

\begin{figure}
	\includegraphics[width=\columnwidth]{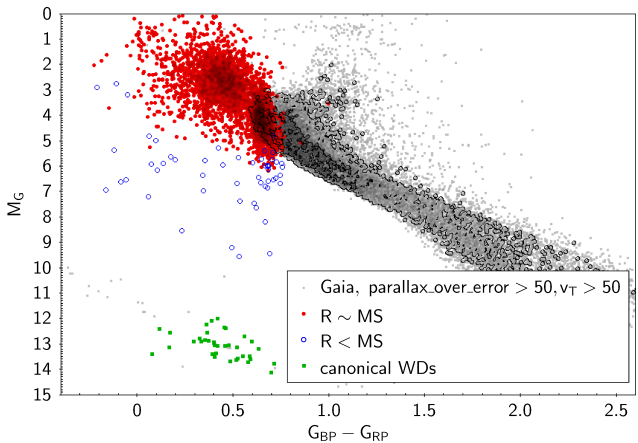}
	\caption{The absolute magnitude $M_G$ as a function of $G_{BP}-G_{RP}$ for {\it Gaia} objects obeying the selection criteria outlined in Section~\ref{filtering}, but with parallax precision better than 2~per cent as well as $v_T > 200$~km/s, is shown in grey. The sdAs with radii consistent with the main sequence are shown in red, those consistent with canonical white dwarfs are shown as green squares, and the objects with intermediate radii are shown as open blue circles. This illustrates that most sdAs seem to be indeed sampled from the halo distribution, given their locus coinciding with the $v_T > 200$~km/s population.}
	\label{haloHR}
\end{figure}

Most of the objects whose radii are not below the main sequence with 99 per cent confidence show a confidence interval spanning both above and below the main sequence lower limit, with only $\approx 20$ per cent showing the 99 per cent confidence level above the main sequence minimum. We cannot make any claims about the nature of individual objects whose radii are not below the ZAMS; they are consistent with both the main sequence and with pre-ELMs, given the occurrence of burning stages that can increase pre-ELM radii above the main sequence, as mentioned above. For example, \citet{liakos2018} recently found a $0.07~M\subsun$ pre-ELM with a $1.0~R\subsun$ radius in an Algol-type binary. However, given the short lifetimes of pre-ELMs \citep[only up to the order of a few million years in the models of][]{istrate2016}, the most likely explanation is that the majority of sdAs are simply metal-poor main sequence stars, as already suggested by \citet{brown2017}. The positions of most of these objects in the color-magnitude diagram of Fig.~\ref{MgBp} are consistent with MIST isochrones \citep{mist1,mist2}, suggesting that they are metal-poor main sequence stars in the halo. Comparison with {\it Gaia} objects showing tangential velocities larger than 200~km/s \citep[a simplified halo selection employed by][]{babusiaux2018} supports this conclusion, as shown in Fig.~\ref{haloHR}. From the isochrones in Fig.~\ref{MgBp} it can be inferred that these main sequence stars show $M < 0.9~M\subsun$ and $6\,000 < T\eff < 8\,000$~K. Such physical parameters are very different compared to those of disc A-type stars ($M \gtrsim 2.0~M\subsun$, $T\eff \sim 10\,000$~K): these objects are not bona-fide A stars, they simply mimic hydrogen-line-dominated A-type spectra due to the low metallicity. They are thus more similar to the classical cool subdwarfs \citep[sdFs and sdGs, e.g.][]{scholz2015}, which are evolutionarily in the main sequence, than to He-burning hot subdwarfs \citep{heber2016}. They likely have masses below 1~$M_\odot$, corresponding to longer main sequence lifetimes that are consistent with the age of the halo. They are also intrinsically fainter than A stars, hence not saturated in the SDSS for distances down to $\sim 2$~kpc, consistent with the halo. A significant number of objects ($\approx 2\,200$), however, lie bluer than the turnoff --- these are possibly blue straggler stars, which are indeed A-type stars and are also believed to be binary byproducts like the ELMs. In this case, the lifetime in the main sequence is extended by mass accretion. A gap between these two populations can be glimpsed in Fig.~\ref{MgBp}, and its position coincides approximately with the turnoff for a metallicity of $[Fe/H] = -1.25$, which is the average metallicity of the analysed sdAs that have SSPP determinations.

\subsection{(Pre-)ELMs in the Galactic disc}

\begin{figure}
	\includegraphics[angle=-90,width=\columnwidth]{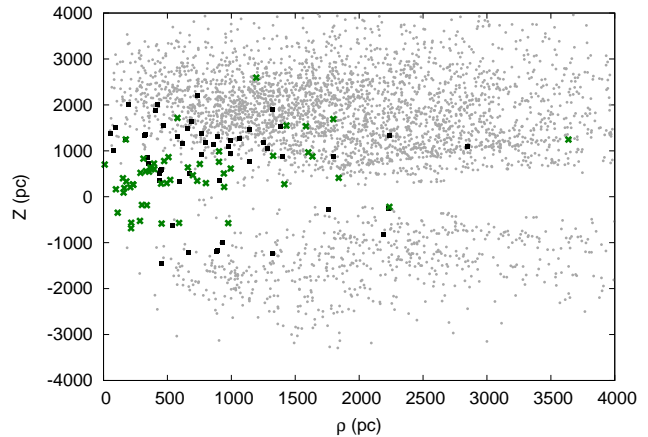}
	\caption{Diagram showing the disc height $Z$ vs. the distance in the Galactic plane $\rho = \sqrt{X^2 + Y^2}$. The sdAs with inconclusive radii are shown in grey, the \mbox{(pre-)ELMs} in the sample are shown as black squares, and the known ELMs from \citet{brown2016} are shown as green crosses. There are more objects at $Z > 0$ and few close to the disk because of the SDSS coverage of the sky.}
	\label{rhoz}
\end{figure}

\begin{figure}
	\includegraphics[angle=-90,width=\columnwidth]{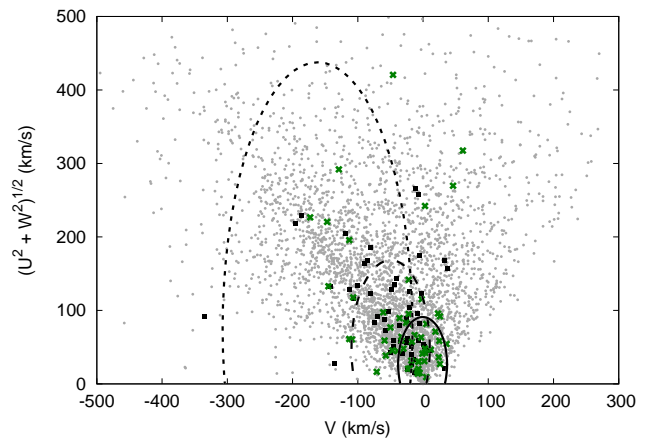}
	\caption{Toomre diagram following same colour coding for the dots as Fig~\ref{rhoz}. The ellipsoids show the 3$\sigma$ contours for Galactic thin disc (solid), thick disc (short-dashed) and stellar halo (long-dashed) populations.} 
	\label{toomre}
\end{figure}

Given their smaller-than-main-sequence radii, the distances of the \mbox{(pre-)ELM} objects are consistent with the Galactic disc, as we had already suggested in \citet{pelisoli2018}. Objects with inconclusive radii extend to larger distances, as expected for a halo population. Fig.~\ref{rhoz} shows a diagram of the disc height $Z$ as a function of $\rho = \sqrt{X^2 + Y^2}$. The \mbox{(pre-)ELMs} are all within 3~kpc. The Toomre diagram (Fig.~\ref{toomre}) also suggests disc membership, with a few objects showing larger velocities, as also happens for the known ELMs. This can be explained by a contribution of the orbital component to the estimated velocities, given that \mbox{(pre-)ELMs} are most likely in close binaries.

\begin{figure}
	\includegraphics[angle=-90,width=\columnwidth]{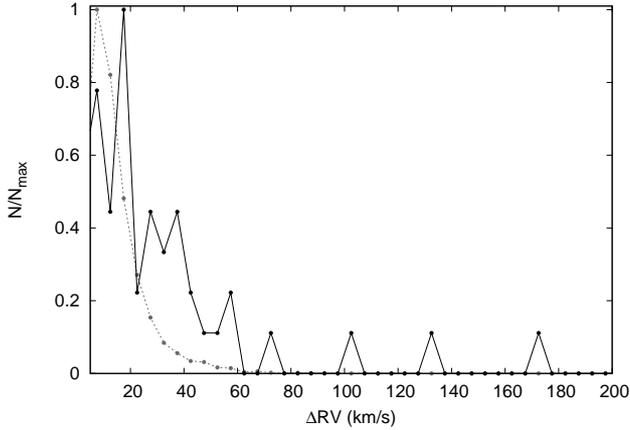}
	\caption{Normalized number density for the maximum difference in radial velocity measured in SDSS subspectra. The black line is for the identified \mbox{(pre-)ELMs}, the dashed grey for the remaining sample. It seems that the \mbox{(pre-)ELMs} show higher velocity amplitudes on average, as expected from a binary evolution origin.}
	\label{histRV}
\end{figure}

\begin{figure}
	\includegraphics[angle=-90,width=\columnwidth]{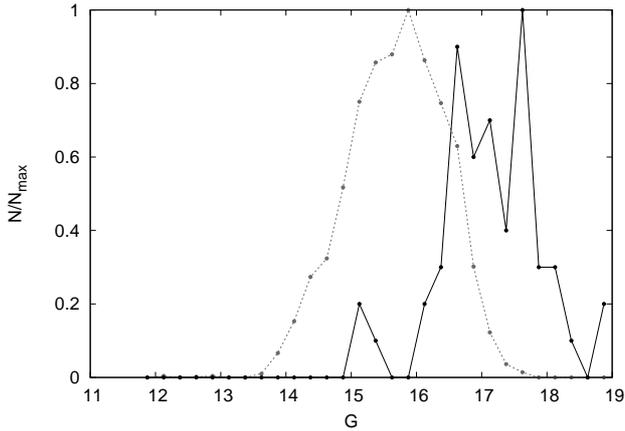}
	\caption{Number of objects as a function of apparent magnitude $G$. The black solid line shows objects whose radii is smaller than main sequence, the grey dashed is for the remaining objects in the sample. The 50 \mbox{(pre-)ELMs} show apparent magnitudes generally fainter than the remaining objects. White dwarfs are intrinsically fainter, so that even at relatively high $G$ magnitudes the parallax can be detected.}
	\label{G}
\end{figure}

Fig.~\ref{histRV} shows the normalized number density of the maximum difference in radial velocity between SDSS subspectra, $\Delta RV$, for \mbox{(pre-)ELMs} compared to the remaining sample. The SDSS time coverage and the signal-to-noise ratio of the subspectra are not enough to claim binarity in most cases \citep{badenes2012}, but it can still be noted that the \mbox{(pre-)ELMs} show more objects with high $\Delta RV$, which is expected considering that they are most likely products of binary evolution. However, it is important to notice that the \mbox{(pre-)ELMs} are mostly fainter than the inconclusive objects (see Fig.~\ref{G}), therefore this could also be due to larger spread resulting from lower $S/N$ spectra. We have analysed only objects whose SDSS final spectra show $S/N > 15$, but the individual subspectra can show significantly lower $S/N$.

\begin{figure}
	\includegraphics[width=\columnwidth]{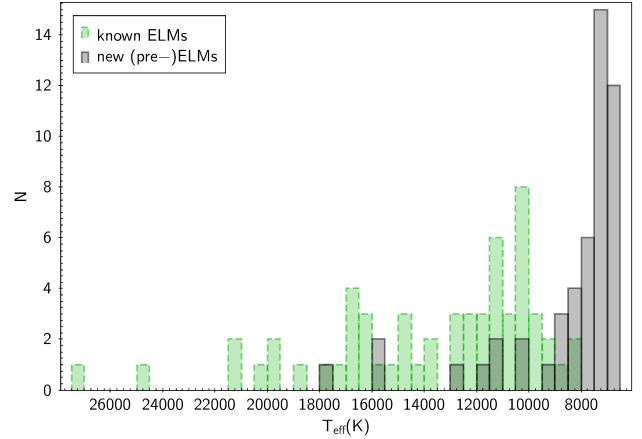}
	\caption{Histogram of $T\eff$ for the objects in Table~\ref{preELMs} (solid, black), compared with the binary ELMs from table 5 of \citet{brown2016} (dashed, green).}
	\label{histT}
\end{figure}

\begin{figure}
	\includegraphics[width=\columnwidth]{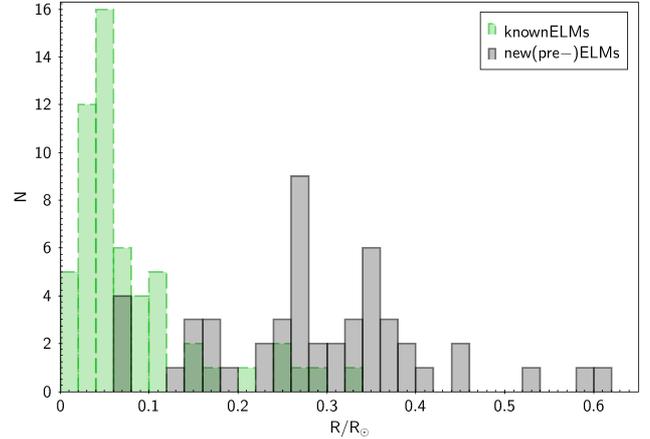}
	\caption{Same as Fig.~\ref{histT} for the radius obtained from the {\it Gaia} parallax.}
	\label{histR}
\end{figure}

Most of the detected \mbox{(pre-)ELMs} show cooler $T\eff$ than the known ELMs of \citet{brown2016}, as shown in the histogram in Fig.~\ref{histT}. This is expected given our selection of A-type spectra, as opposed to the known ELMs that were selected mainly from colours that correspond to hotter B-type stars (see the HR diagram in Fig.~\ref{HR}). Moreover, as our initial sample selection included stars classified as A, B or O stars by the SDSS pipeline, it is biased towards larger \mbox{(pre-)ELMs}, because objects with small radii would most likely be spectroscopically classified as white dwarfs. This is demonstrated in Fig.~\ref{histR}, where we compared the radii of the objects in Table~\ref{preELMs} with the radii we obtain for the objects in the ELM sample.

\section{Summary and Conclusions}

The evolutionary origin of the sdAs has been elusive since they were catalogued by \citet{dr12cat}. It has always been emphasised how {\it Gaia} would be able to shed light on the issue. We analysed the sample of sdAs previously published in \citet{pelisoli2018} in light of {\it Gaia} DR2 data. While parallaxes are unreliable considering our selection criteria for 90 per cent of the sample, we were able to identify 50 new high-probability \mbox{(pre-)ELMs} among the 3\,891 sdAs with reliable {\it Gaia} parallaxes. These identifications increase the number of known \mbox{(pre-)ELMs} by up to 50 per cent. We consider 29 of these new identifications to be most reliably compact objects since they remain below the main sequence when we inflate the uncertainty on parallax to account for systematic uncertainty. The nature of the remaining objects in the sample is still unconfirmed with the {\it Gaia} parallaxes, but their position in the HR diagram supports that most are consistent with a population of low-metallicity main sequence stars in the halo, as first suggested by \citet{brown2017}, with contributions from a low-mass ($M \lesssim 0.9~M\subsun$) population and from a population of blue straggler stars.

The position of the discovered \mbox{(pre-)ELMs} coincides well with the region of the H-R diagram spanned by the binary evolution models of \citet{istrate2016}. This supports a binary evolution origin for the \mbox{(pre-)ELMs}. In fact, we have also noticed that these objects show greater differences in $RV$ measurements from SDSS subspectra compared to the other objects in the sample, supporting that more \mbox{(pre-)ELMs} are in tight binary systems. Yet some show low $RV$ dispersion, hence individual systems cannot be ruled out as single. Although the \mbox{(pre-)ELMs} are more easily explained as having been created in multiple systems, this does not imply that they are currently in close binaries, similarly to the hot subdwarfs sdBs/sdOs \citep[e.g.][]{heber2016}; as mentioned in the Introduction, alternative explanations have been put forward. Time-resolved spectroscopy of the objects in this sample should be obtained so that the present binary ratio of the \mbox{(pre-)ELMs} can be accurately determined. 

As already noted in our previous works \citep{pelisoli2017,pelisoli2018}, the \mbox{(pre-)ELM} radii give the sdAs a distribution of distances consistent with the Galactic disc. The velocities for most objects are consistent with the disc, with a few showing larger velocities, which can possibly be explained by orbital motion. The remaining objects show distances  and velocities that suggest they are sampled from a halo population. Almost a hundred of these show $v_T > 500$~km~s$^{-1}$ and could be high velocity stars.

The new \mbox{(pre-)ELMs} show temperatures cooler than the bulk of known ELMs \citep[which mostly have $T\eff \gtrsim 10\,000$~K; e.g.,][]{elmsurveyVII}. We have previously raised the issue of a missing, cooler population \citep{pelisoli2018}, and pointed out that this missing population was most likely within the sdAs, which seems to be now confirmed.

These identifications might correspond to only a fraction of the \mbox{(pre-)ELMs} within the larger sdA sample, given our conservative choice of minimal main-sequence radius that assumes a very low value of metallicity. The average value of [Fe/H] for the analysed sdAs fitted by SSPP is $-1.33$, with a spread of 0.84, whereas we have assumed a value of [Fe/H] = $-4.0$ to determine the minimum radius threshold. Assuming a higher metallicity would increase the minimum radius limit and the number of objects with radii below it. On the other hand, the number of identifications decreases by 40 per cent when the {\it Gaia} systematics are added to the uncertainty in parallax.

Our work demonstrates the potential of using {\it Gaia} to identify new \mbox{(pre-)ELMs}. Future data releases will further improve our understanding of the sdAs and how they fit into our theories of stellar evolution and Galaxy structure and formation. In particular, {\it Gaia} radial velocities from multiple epochs will enable better constraints on binarity, as would follow-up ground-based spectroscopy. Improved extinction maps would reduce uncertainties in our radius estimates, allowing us to better separate different luminosity classes. Searches independent of the SDSS identifications should be done to provide an unbiased catalogue of \mbox{(pre-)ELMs} \citep[similar to the approach used to identify white dwarfs in {\it Gaia} data by][]{kilic2018, nicola2018}. An unbiased catalogue could help constrain the evolutionary timescales during the pre-ELM phase compared to the cooling-track phases. The predictions from theoretical models are largely affected by uncertainties on residual burning rates. A magnitude-limited sample would also better reveal the Galactic location and kinematics of the sdA subpopulations. Identifications of \mbox{(pre-)ELMs} in the pulsational instability strips can also inform efforts to astroseismically constrain the interiors of post-common-envelope mass-transfer remnants in a regime where spectroscopic classifications have not been reliable \citep{bell2017}. Such studies will not only improve our understanding of the formation of the ELMs themselves, but also of binary evolution as a whole.

\section*{Acknowledgements}

We thank the two anonymous referees for their constructive reports that have helped to substantially improve this manuscript. IP and SOK acknowledge support from CNPq-Brazil. DK received support from programme Science without Borders, MCIT/MEC-Brazil. IP was also supported by Capes-Brazil under grant 88881.134990/2016-01, and by DFG under grant GE 2056-12-1. KJB was supported by the European Research Council under the European Community's Seventh Framework Programme (FP7/2007-2013) / ERC grant agreement no 338251 (StellarAges). We thank Warren Brown, Mukremin Kilic, JJ Hermes, and Saskia Hekker for reading the draft of this work and providing helpful comments, and Alejandra D. Romero for providing single evolution models. We also thank Stefan Jordan, Uli Bastian, and the organizers and participants of the {\it Gaia} Data Workshop held in Heidelberg for useful discussions.

This work has made use of data from the European Space Agency (ESA) mission Gaia (\url{https://www.cosmos.esa.int/gaia}), processed by the Gaia Data Processing and Analysis Consortium (DPAC, \url{https://www.cosmos.esa.int/web/gaia/dpac/consortium}). Funding for the DPAC has been provided by national institutions, in particular the institutions participating in the Gaia Multilateral Agreement. This research has also made use of the TOPCAT (\url{http://www.starlink.ac.uk/topcat/}) software \citep{taylor2005} and of Astropy (\url{http://www.astropy.org}), a community-developed core Python package for Astronomy \citep{astropy:2013, astropy:2018}. 



\bibliographystyle{mnras}
\bibliography{sdAs}



\appendix

\section{Comparison with previous data}
\label{extra}

\begin{figure}
	\includegraphics[angle=-90,width=\columnwidth]{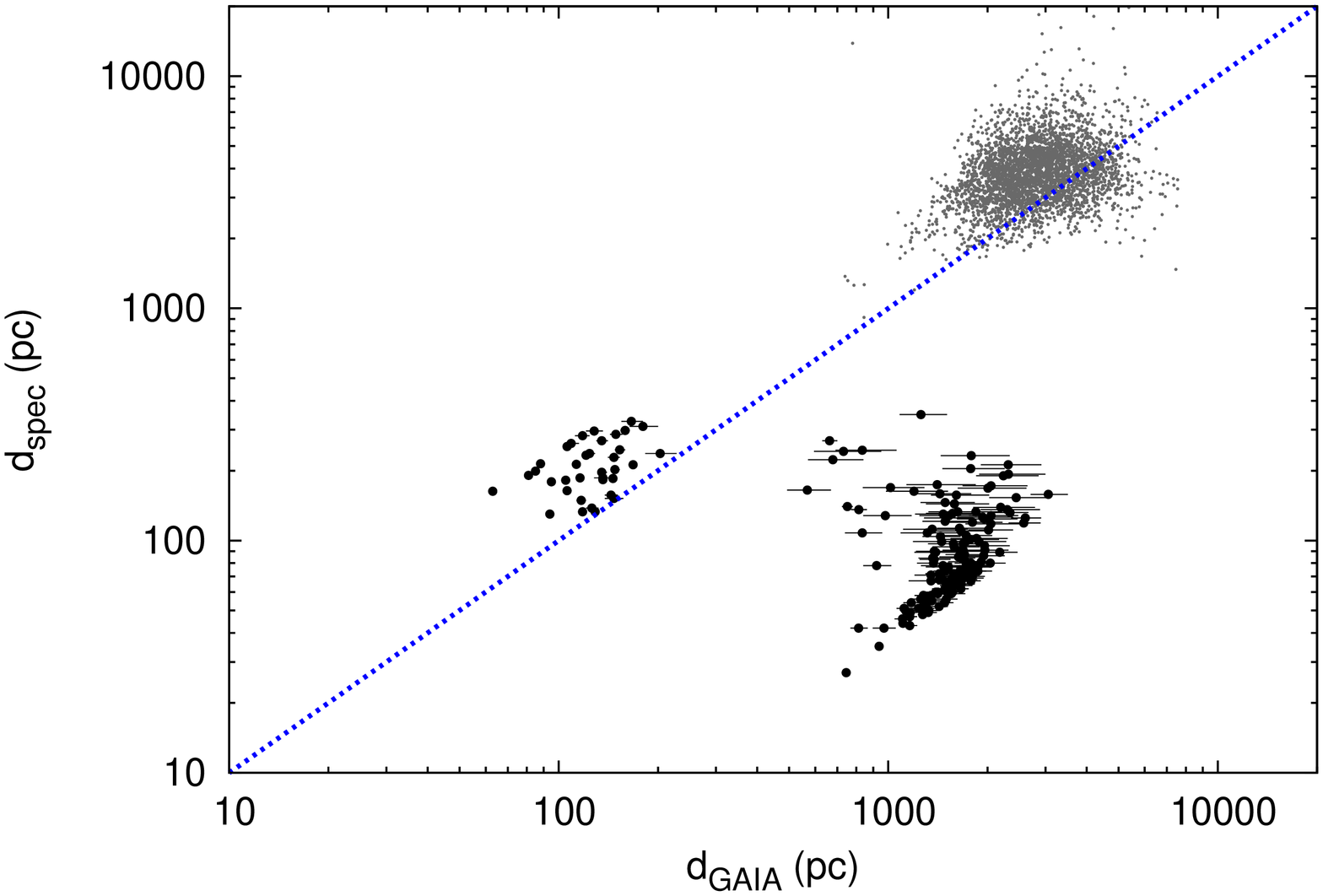}
	\caption{Comparison between the distances estimated from the spectroscopic fits and from {\it Gaia} parallaxes, for the \mbox{(pre-)ELMs} (black) and for the remaining sample (grey). For the \mbox{(pre-)ELMs}, the radii were estimated relying on the $\log~g$, which we found to show a 0.5--1.0~dex uncertainty \citep{pelisoli2018b}, explaining the cloud of objects with distance underestimated by a factor of ten. For the remaining objects, we show the distance obtained from the solid angle, assuming a main sequence radius interpolated given the fitted $T\eff$. There is relatively good agreement with the {\it Gaia} estimate in this case.}
	\label{dist}
\end{figure}

Fig.~\ref{dist} compares the distances obtained from the {\it Gaia} parallaxes with our previous values calculated from the solid angles and radii estimates. For the \mbox{(pre-)ELMs}, we show the distance that we derived given the radii interpolated from the models of \citet{althaus2013} using our spectroscopic $\log~g$. For a significant number of objects, the distance was underestimated by a factor of ten, which is consistent with the systematic uncertainty of $-0.5$ to $-1.0$~dex that we found our $\log~g$ values determined from SDSS spectra to show \citep{pelisoli2018b}. This could also be explained if the radii inferred from the models are smaller than the actual radii, implying an underestimate of the distance. For the remaining objects, the distance was obtained assuming a radius interpolated from solar abundance main sequence models given the $T\eff$. In this case, the agreement seems better, given that our $T\eff$ is reliable to 5 per cent.

\begin{figure}
	\includegraphics[angle=-90,width=\columnwidth]{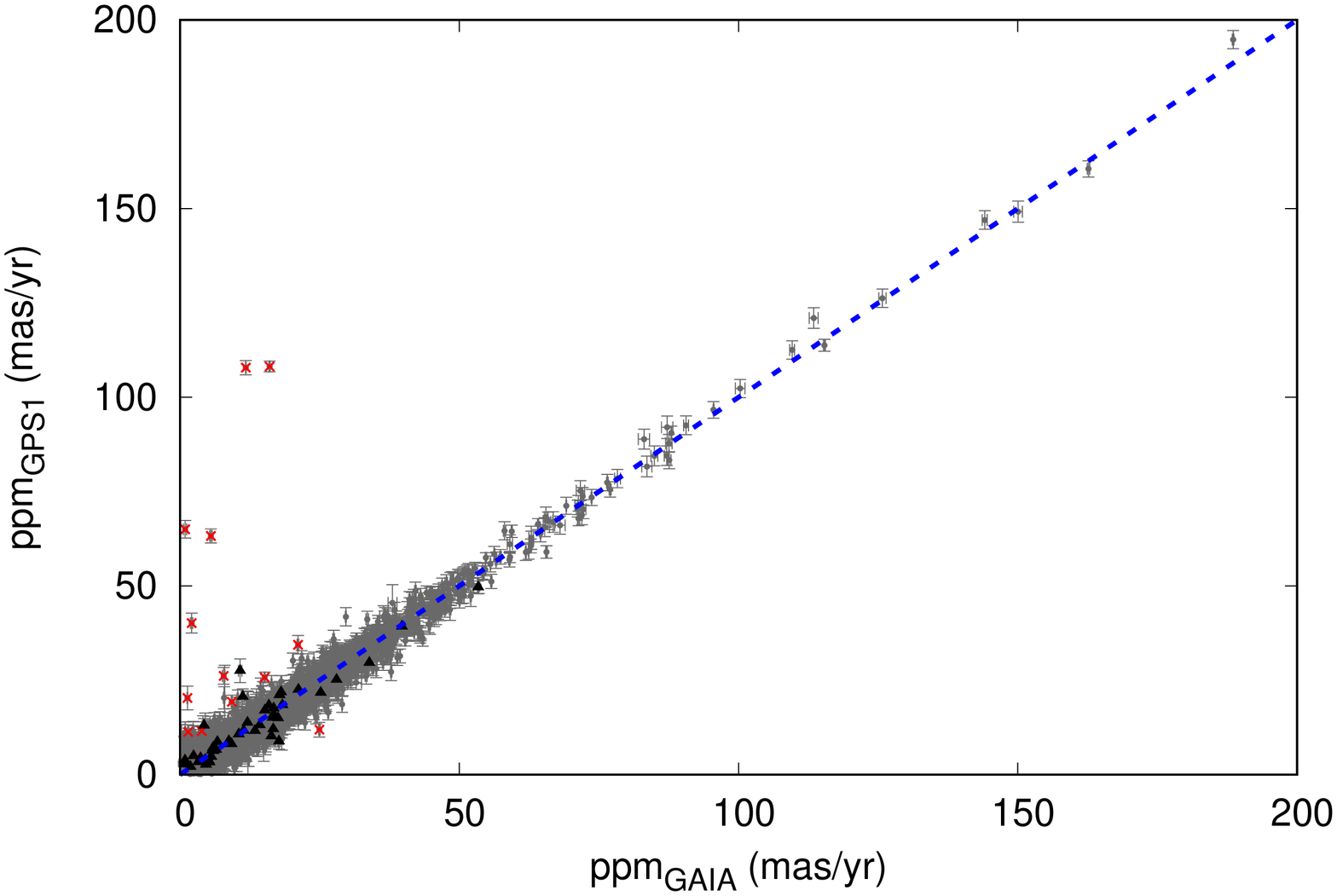}
	\caption{Comparison between the proper motions given in {\it Gaia} and those of the GPS1 catalogue \citep{tian2017}, used in \citet{pelisoli2018}. Objects marked with red crosses show proper motion different between the two catalogues to a $5\sigma$ level. Two per cent of the GPS1 proper motions for this sample are overestimated in a $3\sigma$ level. The objects from Table~\ref{preELMs} are marked in black.}
	\label{ppm}
\end{figure}

\begin{figure}
	\includegraphics[angle=-90,width=\columnwidth]{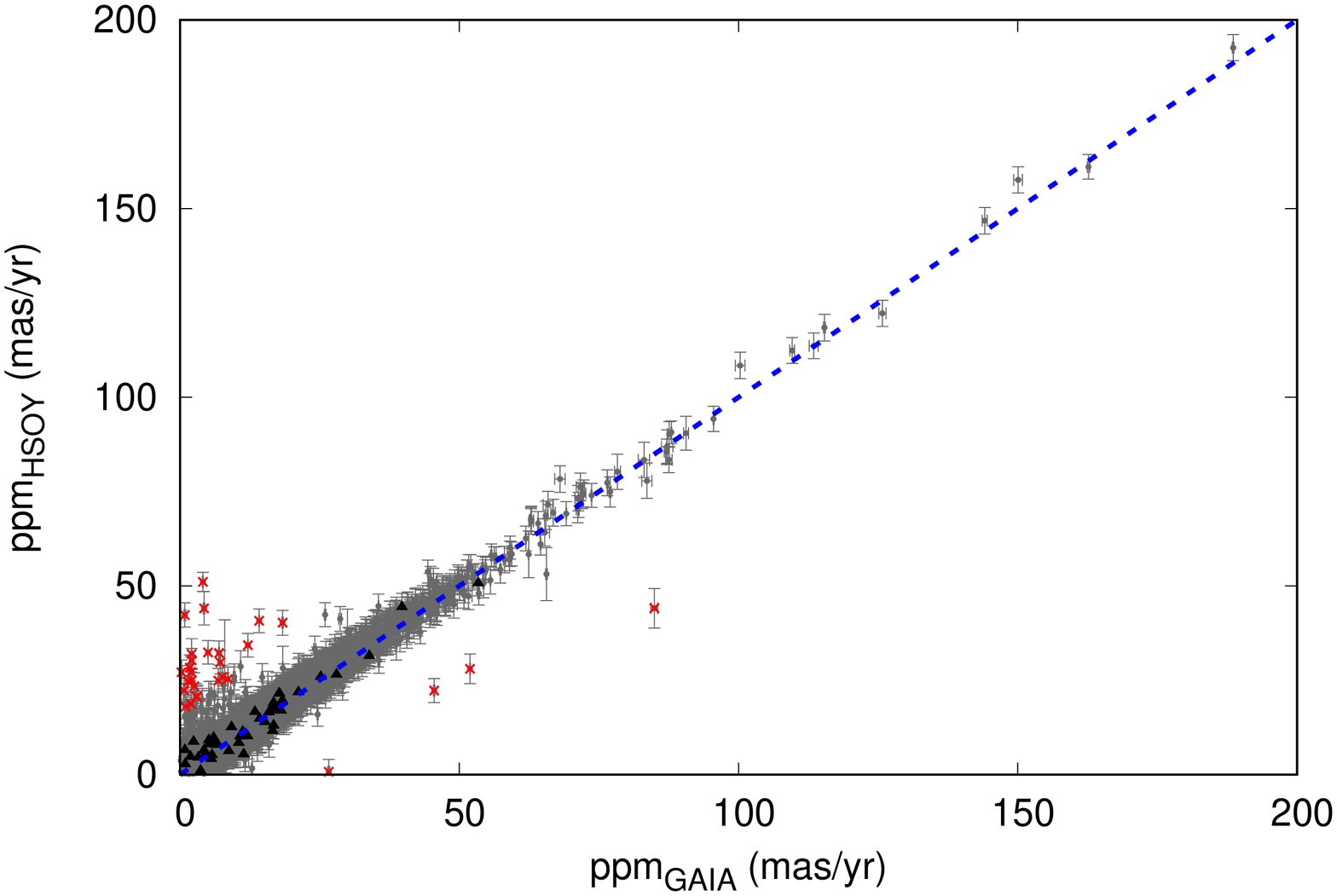}
	\caption{Same as Fig.~\ref{ppm}, for the proper motions of \citet{altmann2017}.}
	\label{ppm2}
\end{figure}

In Fig.~\ref{ppm}, we compare the proper motions used in \citet{pelisoli2018}, obtained from the GPS1 catalogue \citep{tian2017} with the new estimates from {\it Gaia}. The proper motions agree within 3~$\sigma$ for 98 per cent of the sample. A similar result is obtained when comparing with the proper motions of the Hot Stuff for One Year catalogue \cite[HSOY,][]{altmann2017}, as shown in Fig.~\ref{ppm2}. As we relied in proper motions, but not on the distance estimates in \citet{pelisoli2018}, our conclusions there remain unaltered, and are in fact corroborated by this present work: the sdAs are composed of overlapping populations, containing a significant fraction of \mbox{(pre-)ELMs}. In both Figs.~\ref{ppm} and \ref{ppm2}, we mark the 50 objects from Table~\ref{preELMs} to show that the {\it Gaia} proper motion is consistent with previous determinations for these objects, suggesting that the {\it Gaia} solution is correct, as inferred from the filtering parameters discussed in Section~\ref{filtering} and consistent with the fact that these objects are away from dense regions.

\bsp	
\label{lastpage}
\end{document}